\documentclass[12pt]{iopart}

\usepackage{amsgen}
\usepackage{amssymb}
\usepackage{setstack}

\def\be{\begin{equation}}
\def\ee{\end{equation}}
\def\bea{\begin{eqnarray}}
\def\eea{\end{eqnarray}}

\begin{document}

\title{A Gravitational Entropy Proposal}

\author{Timothy Clifton$^1$, George F R Ellis$^{2,3}$ and Reza Tavakol$^1$}
\address{$^1$ School of Physics and Astronomy, Queen Mary University of London.\\
$^2$ Department of Mathematics and Applied Mathematics, University of Cape Town.\\
$^3$ Trinity College and DAMTP, University of Cambridge, Cambridge, UK.}



\begin{abstract}

We propose a thermodynamically motivated measure of gravitational entropy based on the Bel-Robinson tensor, which has a natural interpretation as the effective super-energy-momentum tensor of free gravitational fields. The specific form of this measure differs depending on whether the gravitational field is Coulomb-like or wave-like, and reduces to the Bekenstein-Hawking value when integrated over the interior of a Schwarzschild black hole. For scalar perturbations of a Robertson-Walker geometry we find that the entropy goes like the Hubble weighted anisotropy of the gravitational field, and therefore increases as structure formation occurs. This is in keeping with our expectations for the behaviour of gravitational entropy in cosmology, and provides a thermodynamically motivated arrow of time for cosmological solutions of Einstein's field equations. It is also in keeping with Penrose's Weyl curvature hypothesis.

\end{abstract}

\section{Introduction}
\label{Sec1}

A key question in cosmology is how to define the entropy in gravitational fields. A suitable definition already exists for the important case of stationary black holes \cite{bh}, but in the cosmological setting a well-motivated and universally agreeable analogue has yet to be found.  Addressing this deficit is an important problem, as in the presence of gravitational interactions the usual statements about matter becoming more and more uniform are incorrect. Instead, structure develops spontaneously when gravitational attraction dominates the dynamics \cite{Ell95,Pen11}. This behaviour is crucial to the existence of complex structures, and indeed life, in the Universe. The question then arises, how can evolution under the gravitational interaction be made compatible with the second law of thermodynamics? If the second law is valid in the presence of gravity, such that entropy increases monotonically into the future, then the current state of the universe must be considered more probable than the initial state, even though it is more structured. For this to be true, the gravitational field itself must be carrying entropy.

For a candidate definition of gravitational entropy to be compatible with cosmological processes, such as structure formation in the Universe, it needs to be valid in
non-stationary and non-vacuum spacetimes. We will argue that an appropriate definition of gravitational entropy should only involve the free gravitational field, as specified by the Weyl part of the curvature tensor, $C_{abcd}$ \cite{13a}, and that a particular promising candidate for gravitational entropy can be constructed from integrals of quantities constructed from the pure Weyl form of the Bel-Robinson tensor \cite{bel,robinson}. As we shall see below, this definition has the desired property of increasing as inhomogeneities form through gravitational attraction. It also reduces to the Bekenstein-Hawking value when evaluated in the case of a Schwarzschild black hole.

Situations of particular interest are (i) those with Coulomb-like gravitational fields, representing the relativistic extension of Newtonian gravitational fields; and (ii) those involving wave-like gravitational fields. Each of these can be naturally represented by considering the Petrov classification of the Weyl tensor, and each constitutes a very different type of gravitational interaction. There is no {\it a priori} reason why one should expect the gravitational entropy in these two different settings to be describable in the same way, and we therefore consider them separately in what follows. As we shall see, this separation simplifies the problem of understanding the thermodynamic properties of each of these two types of gravitational fields considerably. The more general case,  in which Coulomb and wave-like parts of the gravitational field are mixed, will, however, require a further and more detailed treatment.

Throughout this paper we will make use of the $1+3$ covariant description of gravitational fields \cite{13a,13b}, which proceeds by taking a timelike unit vector, $u^a$, and defining a projection tensor, $h_{ab} = g_{ab} + u_a u_b$.  The covariant derivative of $u^a$ can then be split into irreducible parts such that
\be
\label{uab}
u_{a;b} = -\dot{u}_a u_b + \frac{1}{3} \Theta h_{ab} +\sigma_{ab} +\omega_{ab},
\ee
where $\Theta = h^{ab} u_{a;b}$ is the expansion scalar, $\sigma_{ab}= (h_{(a}^{\phantom{a} c} h_{b)}^{\phantom{b} d} - \frac{1}{3} h_{ab} h^{cd}) u_{c;d}$ is the shear tensor, $\omega_{ab}=h_{[a}^{\phantom{a} c} h_{b]}^{\phantom{b} d}  u_{c;d}$ is the vorticity tensor, and $\dot{u}^a = u^b u^a_{\phantom{a};b}$ is the acceleration vector. The energy-momentum tensor of a fluid can then be decomposed relative to $u^a$ such that
\be
\label{Tab}
T_{ab}  =\rho u_a u_b + q_a u_b + u_a q_b +p h_{ab} + \pi_{ab},
\ee
where $\rho$ is the energy density, $q^a$ is the momentum density, $p$ is the isotropic pressure, and $\pi_{ab}$ is the anisotropic (tracefree) pressure of the fluid. In analogy to the decomposition of the Maxwell tensor into electric and magnetic parts, the Weyl tensor can also be decomposed into electric and magnetic parts as
\be
E_{ab} = C_{abcd} u^c u^d \qquad {\rm and} \qquad
H_{ab} = \frac{1}{2} \eta_{acd} C^{cd}_{\phantom{cd} be} u^e,
\ee
where $\eta_{abc} = \eta_{abcd} u^d$ is the spatial alternating tensor ($\eta_{abcd}=\eta_{[abcd]}$, $\eta_{0123}=\sqrt{\vert g_{ab} \vert}$), and both $E_{ab}$ and $H_{ab}$ are symmetric, tracefree and orthogonal to $u^a$.  We will also make use of the complex null tetrad
\be
\label{nullt}
m^a = \frac{1}{\sqrt{2}} \left( x^a - i y^a \right), \quad
l^a = \frac{1}{\sqrt{2}} \left( u^a - z^a \right), \quad {\rm and} \quad
k^a = \frac{1}{\sqrt{2}} \left( u^a + z^a \right),
\ee
where $x^a$, $y^a$ and $z^a$ are the spacelike unit vectors that, together with $u^a$, form an orthonormal basis, and $g_{ab} =2 m_{(a} \bar{m}_{b)} - 2 k_{(a} l_{b)}$.

In Section \ref{Sec2} we discuss our requirements for a sensible definition of gravitational
entropy. These include
being non-negative, vanishing when the Weyl tensor vanishes,
being a function of the anisotropy in the gravitational field, reproducing existing
results for black hole entropy, and increasing monotonically in cosmological solutions
when structure forms. In Section \ref{sec:BR} we introduce and discuss the properties of the
Bel-Robinson tensor, which plays a central role in our proposal.  Section \ref{Sec4} constructs a definition of
entropy from the Bel-Robinson tensor, and from semi-classical notions of the temperature of a gravitational field.
In Section \ref{Sec5} we apply our definition of entropy to the case of a Schwarzschild black hole, and find that
it reproduces the Bekenstein-Hawking value. Section \ref{lss} then contains an application
of our measure to scalar perturbations
about a Robertson-Walker geometry, while in Section \ref{Sec7} we use the example of exact
Lema\^{i}tre-Tolman-Bondi solutions to demonstrate the applicability of our proposed measure
to non-perturbative inhomogeneous cosmological settings. Finally, in Section \ref{Sec8},
we discuss our results.

\section{Requirements for Gravitational Entropy}
\label{Sec2}

In general relativity the gravitational field can be split into Ricci and Weyl parts. The former is related pointwise
to the energy-momentum tensor, where standard definitions for the entropy of matter fields should hold. Counting the
entropy in the Ricci curvature of space-time would therefore be like counting the entropy in the matter fields twice.
As our aim is to characterise the gravitational entropy of free gravitational fields, we therefore choose to concentrate on the Weyl part of the curvature tensor. This provides us with a tensorial
description of the free part of the gravitational field,
and is present even in the absense of matter fields.

Definitions of gravitational entropy constructed from the Weyl tensor have been considered before, including the study of the simple choice $S=C_{ab}^{\phantom{ab} cd} C_{cd}^{\phantom{cd} ab}$. Unfortunately, this form of $S$ has been argued to fail at isotropic singularities \cite{Wainwright-etal}, and also fails to handle the decaying and growing perturbation modes \cite{Rothman-etal}. Another choice has been the dimensionless scalar $S=C_{ab}^{\phantom{ab} cd} C_{cd}^{\phantom{cd} ab}/R_a^{\phantom{a} b} R_b^{\phantom{b} a} $, which while addressing the previous objections does not seem to give the correct sense of time for a radiating source \cite{Bonnor},
and also diverges for the vacuum case. 
Another shortcoming of these earlier definitions is that they are
not always non-negative, and therefore cannot guarantee the monotonicity that is required of an entropy measure.
They also appear as somewhat {\it ad hoc}, in that while they are motivated by the appropriate inclusion of the Weyl tensor in the numerator, they do not make any contact with either thermodynamics or information theory. The latter definition is also dimensionless, meaning it cannot reproduce the Bekenstein-Hawking result when applied to a black hole.

To make progress on this issue we therefore make the following list of requirements on the
gravitational entropy, $S_{\rm grav}$, that we expect to guide us:
\begin{itemize}
  \item \emph{\textbf{E1}}: It should be non-negative.
  \item \emph{\textbf{E2}}: It should vanish if and only if $C_{abcd}=0$.
  \item \emph{\textbf{E3}}: It should measure the local anisotropy of the free gravitational field.
  \item \emph{\textbf{E4}}: It should reproduce the Bekenstein-Hawking entropy of a black hole.
  \item \emph{\textbf{E5}}: It should increase monotonically, as structure forms in the universe.
\end{itemize}
As is the case for energy densities and pressures, we expect a sensible definition of gravitational entropy to be observer dependent (although it should be able to be defined covariantly). While we are not considering quantum gravity here, it will also be considered beneficial if our definition of gravitational entropy can be linked to semi-classical calculations, in a similar way to the link between the Bekenstein-Hawking entropy and Hawking radiation. Finally, we note that we expect the entropy in gravitational and matter fields to be additive, such that the total of entropy in all fields is an extrinsic quantity.

\section{The Bel-Robinson Tensor}
\label{sec:BR}

With the above requirements in mind, we base our measure on the Bel-Robinson tensor, which is defined in
terms of the Weyl tensor as \cite{bel,robinson}
\begin{equation}\label{def2}
T_{a b c d} \equiv \frac{1}{4} \left( C_{e a b f}C^{e \; \; \; \; f}_{\; \; c d \;
\;}+C^{*}_{\; e a b f}C^{* \; e \; \; \; \; f}_{\; \; \; \; \; c d
\; \;} \right),
\end{equation}
where $C^{*}_{\; abcd}= \frac{1}{2} \eta_{abef} C^{ef}_{\phantom{ef}cd}$ is the dual of the Weyl tensor. This tensor is overall symmetric, tracefree, and is covariantly conserved in vacuum (or in the presence of $\Lambda$). The factor of $1/4$ here is included to give a natural interpretation of the Bel-Robinson tensor in terms of the Weyl spinor \cite{PR}. A measure of gravitational entropy constructed from this tensor has already been considered by Pelavas and Lake \cite{Pelavas-Lake} and Pelavas and Coley \cite{coley} in the form $S=\int W d\tau$, where
\begin{equation}\label{def1}
{W} = {T}_{a b c d} u^a u^b u^c u^d.
\end{equation}
One can note immediately that $W$ has the properties of being observer dependent and non-negative. 
We argue that observer dependence (i.e. dependence on $u^a$) is to be expected: this is the case for the energy density $\rho$ and pressure $p$ \cite{13b,13a}, and so may be expected also in this case.
It also vanishes if and only if the Weyl tensor vanishes. This immediately addresses points {\bf E1} and {\bf E2}. Similarly point {\bf E3} can be seen to be satisfied for perfect fluids, as the constraint equation for $E_{ab}$ in this case can be written
\begin{equation}\label{divE}
D^bE_{ab}= - 3\omega^b H_{ab} + \frac{1}{3}D_a\rho + [\sigma,H]_a ,
\end{equation}
where $D_a=h_a^{\phantom{a} b} \nabla_b$ is the orthogonally projected 3-dimensional covariant derivative and $[X,Y]_a=\eta_{abc} X^b_{\phantom{b}d} Y^{cd}$ is a commutator \cite{MaaBas97}. Inhomogeneity in $\rho$ then tells us that either $E_{ab}$ or $H_{ab}$ must be non-zero, and so we must have $W>0$. Hence, inhomogeneity requires both anisotropy and
a non-zero $W$. This is supported by the result that a conformally flat barotropic fluid solution is a Robertson-Walker geometry \cite{Ell71}, which implies that $W=0$ if and only if the spacetime is homogeneous and isotropic at every point.

Rather than constructing our entropy measure as an integral along a timelike curve, we shall employ
integrals over spacelike hypersurfaces (see \cite{hyper}). We also wish to make use of the gravito-electromagnetic properties of the Weyl tensor, as explored by Maartens and Bassett \cite{MaaBas97}. This starts with the recognition that the Bel-Robinson tensor given in Eq. (\ref{def2}) is the unique Maxwellian tensor \cite{max} that can be constructed from the Weyl tensor, and that it acts as the ``super-energy-momentum'' tensor for gravitational fields \cite{sup1,sup2}. This can be seen by considering the $1+3$ decomposition of the relevant evolution and constraint equations, which in vacuum are given by \cite{bel,MaaBas97}
\begin{eqnarray}
&{W} = \frac{1}{4} \left( E_{a}^{\phantom{a} b} E^{a}_{\phantom{a} b} + H_{a}^{\phantom{a} b} H^{a}_{\phantom{a} b} \right) \\
&J_a =\frac{1}{2} [E,H]_a\\
&\dot{W} + D^a J_a \simeq 0,
\end{eqnarray}
where  $J_a = - h_a^{\phantom{a} e} T_{e b c d} u^b u^c u^d$ is the ``super-Poynting vector" and $W$ is the ``super-energy density", which can both be seen to arise naturally as invariants under spatial duality rotations in direct analogy with the energy density and Poynting vector of Maxwell's theory \cite{MaaBas97}. The $\simeq$ here indicates that terms that are expected to be small in the weak field limit have been discarded (for the full relativistic equation see \cite{MaaBas97}). The Bel-Robinson tensor therefore behaves like an effective energy-momentum tensor for the free gravitational fields, but has dimensions of $L^{-4}$, rather than the usual $L^{-2}$ (where $L$ is the unit of length). To find effective energy densities and pressures with the correct dimensionality we must therefore take a square-root. This problem has been considered by Bonilla and Senovilla \cite{bon}, as we will now explain.

For a symmetric and tracefree four-index tensor, such as $T_{abcd}$, one can define a symmetric two-index ``square-root'', $t_{ab}$, as a solution of the following equation \cite{Schouten}:
\be
\label{sr}
T_{abcd} = t_{(a b} t_{cd)} - \frac{1}{2} t_{e (a} t_{b}^{\phantom{b} e} g_{c d)} - \frac{1}{4} t_{e}^{\phantom{e} e} t_{(a b} g_{cd)} +\frac{1}{24} \left( t_{ef} t^{ef} +\frac{1}{2} (t_{e}^{\phantom{e}e})^2 \right) g_{(ab} g_{cd)}.
\ee
The right-hand side of this equation is the only totally symmetric and tracefree four-index tensor that be constructed that is quadratic $t_{ab}$. While it can be verified that for a given $t_{ab}$ this equation gives the only possible symmetric and tracefree four-index tensor, it is not the case that there exists a $t_{ab}$ for any arbitrary symmetric and tracefree four-index tensor. Furthermore, it can be seen that for any solution, $t_{ab}$, there also exists another solution $\epsilon t_{ab} + f \; g_{ab}$, where $\epsilon=\pm1$ and where $f$ is an arbitrary function \cite{bon}. We will explain in what follows how these ambiguities can be removed from our considerations for the cases that interest us here.

The solution to Eq. (\ref{sr}) exists and is unique for tracefree $t_{ab}$ (i.e. when $f=0$) in spacetimes of Petrov type D or N \cite{bon}. The tracefree requirement does not, however, necessarily lead to a quantity that is conserved in vacuum. One can therefore choose for the square-root of the Bel-Robinson to inherit its tracefree property, or (at least part of) its property of conservation in vacuum, but not necessarily both simultaneously. For the case of Petrov type D, in which there exist two double principal null directions, and the gravitational field is Coulomb-like, the tracefree square-root can be written as \cite{bon}
\be
\label{em}
t_{ab} = 3 \epsilon \vert \Psi_2 \vert \left( m_{(a} \bar{m}_{b)} + l_{(a} k_{b)} \right),
\ee
where $\Psi_2= C_{abcd} k^a m^b \bar{m}^c l^d$ is the only non-zero Weyl scalar, and where we have used a complex null tetrad as defined in Eq. (\ref{nullt}) with $l^a$ and $k^a$ aligned with the two principal null directions. For Petrov type N spacetimes, in which all four principal null directions are degenerate, and the gravitational field is wave-like, the tracefree square-root can be written as \cite{bon}
\be
\label{em2}
t_{ab} = \epsilon \vert \Psi_4 \vert k_a k_b,
\ee
where $\Psi_4 = C_{abcd} \bar{m}^a l^b \bar{m}^c l^d$ is the only non-zero Weyl scalar in this case, and $k^a$ is chosen to be aligned with the principal null directions. This is obviously very similar to the energy-momentum tensor of pure radiation, and serves to further substantiate the claim that these objects behave like the effective energy-momentum of free gravitational fields.

For other Petrov types the Bel-Robinson tensor can be factored into terms that are either Coulomb-like (as in the type D result above), wave-like (as in the case of type N), or more complicated. While the tracefree square-roots given above are unique, the factorisation of the Bel-Robinson tensor into two different symmetric tracefree tensors is not in general. This suggests that spacetimes for which the free gravitational fields are a mixture of wave-like and Coulomb-like parts could be more complicated, and that defining an effective energy density in free gravitational fields in these cases could also be complicated. Nevertheless, the cases given above contain very interesting examples for us to consider. These include all stationary black hole solutions, which are useful to compare to established definitions of gravitational thermodynamics, as well as the case of scalar perturbations to Robertson-Walker geometries, which are of great interest for cosmology. It is also the case that in all spacetimes with non-zero Weyl curvature a factor of the Bel-Robinson tensor exists which takes the form given in either Eq. (\ref{em}) or (\ref{em2}). This motivates using these expressions to determine the effective energy-momentum tensor for the Coulomb-like and wave-like parts of the free gravitational field, respectively.

\section{A Thermodynamic Description of the Free Gravitational Field}
\label{Sec4}

In constructing our definition of gravitational entropy we will be seeking to construct a gravitational analogue of the fundamental law of thermodynamics
in the form:
\be
\label{first}
T_{\rm grav} d S_{\rm grav} =  dU_{\rm grav} + p_{\rm grav} dV,
\ee
where $T_{\rm grav}$, $S_{\rm grav}$, $U_{\rm grav}$ and $p_{\rm grav}$ denote the effective temperature, entropy, internal energy and isotropic pressure of the free gravitational field, respectively, and $V$ is the spatial volume. This will require a brief consideration of relativistic thermodynamics, as derived from the energy-momentum conservation equations, and the use of the square-root of the Bel-Robinson tensor to create an effective energy-momentum tensor for the free gravitational field\footnote{
Please note that by defining quantities such as $U_{\rm grav}$ and $p_{\rm grav}$ we do not intend to imply that the free gravitational field should contribute to the right-hand side of the Einstein
equations, only that quantities can be identified that appear to obey equations that are
closely analogous to those of actual matter fields, and that these equations can be used to construct a definition of gravitational entropy that has many of the properties we require.}.

\subsection{Relativistic Thermodynamics}

The subject of relativistic thermodynamics has been considered by a number of authors in the past
(see, e.g., \cite{thermrev} for a review). Starting from the equation for energy-momentum conservation we
immediately find
\be
\label{uT}
(u_a T^{ab})_{;b} = u_{a;b} T^{ab} - u_a J^a,
\ee
where $J^a= -T^{ab}_{\phantom{ab} ;b}$, such that $u_a J^a$ is heat flow into the fluid.
If we now define $\Theta = \dot{v}/v$, then Eqs. (\ref{uab}), (\ref{Tab}) and (\ref{uT}) tell us that
\be
\label{gibbs}
(\rho v)\dot{\;} +p \dot{v} = v \left[ u_a J^a - q^b_{\phantom{b};b} - \dot{u}^a q_a -\sigma_{ab} \pi^{ab}\right],
\ee
where dots denote time derivatives along $u^a$. This looks very much like a relativistic version of the fundamental
thermodynamic equation, (\ref{first}). Given its form, it therefore seems natural to identify
the left-hand side of this equation with $T \dot{s}$, where $s$ and $T$ are the point-wise entropy of the fields described by $T^{ab}$, and their temperature, respectively. That is, we define
\be
T \dot{s} \equiv (\rho v)\dot{\;} +p \dot{v},
\ee
where the entropy in a spatial domain $\sigma$ with volume $V=\int_{\sigma} v$ will be given by $S=\int_{\sigma} s$. Clearly, in order to find $S$ one first needs to define $T$ independently. We shall return to this in Sec. \ref{Temp}, below.

\subsection{Effective Energy-Momentum of Coulomb-like Gravitational Fields}

Following the discussion in Section \ref{sec:BR}, we take the effective energy-momentum of the Coulomb-like gravitational fields that are present in a Petrov type D spacetime, $\mathcal{T}_{ab}$, to be given by the solution to Eq. (\ref{sr}), with a traceless part prescribed by Eq. (\ref{em}), such that
\bea
\label{TtypeD}
8\pi  \mathcal{T}_{ab} &=& \alpha\left[ 3 \epsilon \vert \Psi_2 \vert \left( m_{(a} \bar{m}_{b)} + l_{(a} k_{b)} \right) + f g_{ab} \right]\\
&=& \alpha \left[ \left( \frac{3}{2} \epsilon \vert \Psi_2 \vert +f \right) \left( x_a x_b +y_a y_b\right) - \left( \frac{3}{2} \epsilon \vert \Psi_2 \vert - f \right)\left(  z_a z_b -u^a u^b \right) \right], \nonumber
\eea
where $\alpha$ is a constant to be determined, and in going to the second line we have transformed to a set of orthonormal basis vectors. We have used a curly $\mathcal{T}_{ab}$ here to distinguish this quantity from the actual energy-momentum tensor of matter fields (as would appear on the right-hand side of Einstein's equations).

We now contract this effective energy-momentum tensor with the unit 4-vector, $u^a$, and the projection tensor, $h_{ab}$, to gain the effective energy density, pressure and momentum density. These are given by
\bea
8\pi \rho_{\rm grav} &=& \alpha \left( \frac{3}{2} \epsilon \vert \Psi_2 \vert -f \right), \qquad  \qquad
8\pi p_{\rm grav} = \alpha\left( \frac{1}{2} \epsilon \vert \Psi_2 \vert +f \right)\\
8\pi \pi^{\rm grav}_{ab} &=& \frac{\alpha}{2} \epsilon \vert \Psi_2 \vert  \left( x_a x_b +y_a y_b -z_a z_b + u^a u^b \right),
\label{piD}
\eea
and $q^{\rm grav}_a = 0$. The presence of the free function $f$ in these expression is clearly undesirable, and can be removed by imposing the condition of energy conservation, $u_a \mathcal{T}^{ab}_{\phantom{ab};b}=0$, in vacuum. As mentioned above, such a condition does not follow directly from the vacuum equation $T^{abcd}_{\phantom{abcd};d}=0$ and Eq. (\ref{sr}), but must be imposed separately. In fact, we find it is satisfied if and only if
\be
f=-\frac{1}{2} \epsilon \vert \Psi_2 \vert +\lambda_1,
\ee
where $\lambda_1$ is an arbitrary constant that we will set to zero, as it does not affect the construction of the thermodynamic quantities that interest us. This result is found by differentiating Eq. (\ref{TtypeD}), and using $\vert \Psi_2 \vert = \sqrt{2 W/3}$ together with Eq. (45) from \cite{MaaBas97}. The effective energy-momentum tensor is then given by
\be
\label{TtypeD2}
8 \pi \mathcal{T}_{ab} =  \epsilon \alpha \sqrt{\frac{2 W}{3}} \left( x_a x_b +y_a y_b\right) - 2\epsilon \alpha \sqrt{\frac{2 W}{3}}  \left(  z_a z_b -u^a u^b \right),
\ee
and the effective energy density and pressure in the free gravitational field become
\be
\label{rho}
8\pi \rho_{\rm grav} = 2 \alpha \sqrt{\frac{2 W}{3}}    \qquad  {\rm and} \qquad
p_{\rm grav} =0,
\ee
where we have set $\epsilon =+1$, so that $\rho_{\rm grav} \geq 0$. The anisotropic pressure and momentum density are unchanged from Eq. (\ref{piD}) and $q^{\rm grav}_a=0$, as $f$ only occurs in the trace of $\mathcal{T}_{ab}$. These are the unique expressions for effective energy density and pressure of free gravitational fields in Petrov type D spacetimes that can be obtained from the square-root of the Bel-Robinson tensor under the condition of energy conservation in vacuum and positive energy density\footnote{Up to the possible inclusion of $\lambda_1$, which would act like an effective vacuum term in Eq. (\ref{TtypeD2}).}.

We can now construct from $\mathcal{T}_{ab}$ in Eq. (\ref{TtypeD2}) an effective fundamental
thermodynamic equation, of the form given in Eq. (\ref{gibbs}).
In the presence of perfect fluid matter fields this is
\be
\label{gibbsD}
T_{\rm grav} \dot{s}_{\rm grav} = (\rho_{\rm grav} v)\dot{\;} = -v \sigma_{ab} \left[ \pi_{\rm grav}^{ab}+ \frac{4 \pi (\rho+p)}{3 \alpha \sqrt{2W/3}}  E^{ab}  \right].
\ee
The first term in brackets on the right-hand side of this equation can be seen to be the direct analogue of the relativistic discipation that is present in Eq. (\ref{gibbs}), while the second term is due to the ``heat flow'' into the free gravitational fields from the matter fields. This term vanishes when $\rho=p=0$, or when $p=-\rho$, or $\sigma_{ab}=0$ or $E_{ab}$ =0. While behaving as the analogue of heat in this equation, one should again bear in mind that the energy-momentum of the matter fields is being conserved as usual. This effective heat flow does not therefore signify any exchange of actual energy, but can be used in order to define entropy in the free gravitational fields as we have just done.

To proceed further we need to quantify what we mean by gravitational temperature, $T_{\rm grav}$,
in these settings. This we shall do in Sec. \ref{Temp}.
Finally, let us note that the result $\vert \Psi_2 \vert = \sqrt{2 W/3}$ means that the interesting properties of the Bel-Robinson tensor are inherited by our effective energy-momentum tensor $\mathcal{T}_{ab}$. This includes the satisfaction of points {\bf E1}, {\bf E2} and {\bf E3}, as discussed in Section \ref{sec:BR}.

\subsection{Effective Energy-Momentum of Wave-like Gravitational Fields}

Let us now consider plane-fronted transverse gravitational waves (of the sort that LIGO, and other gravitational wave experiments, are trying to detect). The geometry associated with these waves belongs to Kundt's class \cite{kundt}, which are the class of all Petrov type N solutions with vanishing Newman-Penrose scalar $\rho_{\rm NP}=-k_{a;b}m^a \bar{m}^b$. These types of waves are closely analogous to our understanding of electromagnetic waves, and fit in with the idea of gravitational wave fronts as the congruence of null curves they follow is irrotational.

Taking the effective energy-momentum tensor of this type of gravitational field to be given by the solution to Eq. (\ref{sr}), with a traceless part prescribed by Eq. (\ref{em2}), leads to the effective energy-momentum tensor for the free gravitational fields being of the form
\be
\label{GWT}
8\pi \mathcal{T}_{ab} = \beta\left[ \epsilon \vert \Psi_4 \vert k_a k_b + f g_{ab} \right]
\ee
where $\beta$ is a constant to be determined (which may or may not be equal to $\alpha$ in Eq. (\ref{TtypeD})). In order to have energy conservation in vacuum in this case, such that $u_a \mathcal{T}^{ab}_{\phantom{ab} ;b} =0$, we find that it is sufficient to take $f=\lambda_2=$constant. That is, in this particular case, the tracefree condition and the energy conservation condition can both be satisfied simultaneously.

The effective energy density, pressure, and momentum density of this effective fluid are then given by
\bea
8\pi \rho_{\rm grav} &=& \beta \left( \frac{\epsilon}{2} \vert \Psi_4 \vert -\lambda_2 \right)\\
8\pi p_{\rm grav} &=& \beta \left( \frac{\epsilon}{6} \vert \Psi_4 \vert + \lambda_2 \right)\\
8\pi \pi_{ab}^{\rm grav} &=& - \beta \left( \frac{\epsilon}{6} \vert \Psi_4 \vert \left( x_a x_b + y_a y_b - 2 z_a z_b \right)\right)\\
8\pi q^{\rm grav}_{a} &=& \beta \frac{\epsilon}{2} \vert \Psi_4 \vert z_a,
\eea
and in the expressions for $\pi_{ab}^{\rm grav}$ and $q^{\rm grav}_{a}$ we have used a set of orthonormal basis vectors. We can now make the identification $\vert \Psi_4 \vert = \sqrt{4 W}$, and can consistently and without loss of generality set $\lambda_2=0$ (as it is only the derivative of $f$ that appears in the fundamental thermodynamic equation (\ref{gibbs})). Our thermodynamic quantities for the wave-like gravitational fields being considered in this section then become
\be
8 \pi \rho_{\rm grav} = \beta \sqrt{4 W}  \qquad  {\rm and} \qquad
p_{\rm grav} = \frac{1}{3} \rho_{\rm grav},
\ee
together with $\pi_{ab}^{\rm grav} = -\beta \sqrt{4 W} \left( x_a x_b + y_a y_b - 2 z_a z_b \right)/48 \pi$ and $8 \pi q^{\rm grav}_{a} = \beta \sqrt{4W} z_a$, where we have taken $\epsilon=+1$ so that $\rho_{\rm grav}\geq 0$. The effective fluid in this case therefore takes the form of radiation-like matter fields, with an equations of state $w=p/\rho=1/3$. This can be compared with the dust-like equation of state that occurred in the case of Coulomb-like gravitational fields considered above.

Once again, from these considerations we can construct an effective gravitational analogue of the
thermodynamic equation, (\ref{gibbs}). In the presence of a perfect fluid this is
\bea
\label{gibbsN}
T_{\rm grav} \dot{s}_{\rm grav}  &=& (\rho_{\rm grav} v)\dot{\;} +p_{\rm grav} \dot{v} \nonumber\\
&=& - v \left[ g^{ab} q^{\rm grav}_{a;b} + \dot{u}^a q^{\rm grav}_a + \sigma^{ab} \pi^{\rm grav}_{ab}\right] - \frac{2 \pi v (\rho+p) \sigma_{ab} E^{ab}}{\beta \sqrt{4 W}} .
\eea
Once more, the first term in brackets on the last line of this equation can be seen to be the direct analogue of the relativistic dissipation that is present in Eq. (\ref{gibbs}), while the second term is due to the ``heat flow'' into the free gravitational fields from the matter fields. This term again vanishes when $\rho=p=0$, or $p=-\rho$, or when $\sigma_{ab}=0$ or $E_{ab}$ =0. For the case of general Petrov type N  spacetimes there is an additional term on the right-hand side of this equation of the form
\be
+\frac{4\pi}{\sqrt{2} \beta} \sqrt{4 W} v k_{a;b} (x^a x^b + y^ay^b).
\ee
While this term vanishes for the plane wave geometries in Kundt's class of solutions, it can be non-zero in general \cite{rt}.
Finally, we note that the result $\vert \Psi_4 \vert = \sqrt{4W}$ again means that points {\bf E1}, {\bf E2} and {\bf E3} are satisfied.

As an example of the energy-momentum of a type N solution one could consider an exact plane-fronted gravitational wave with line-element \cite{gw1,gw2}
\be
ds^2 = - dw dv + L^2 \left[ e^{2 \gamma} dx^2 + e^{-2 \gamma} dy^2 \right],
\ee
where $L=L(w)$ and $\gamma = \gamma (w)$. The vacuum field equations then reduce to $L^{\prime \prime} + (\gamma^{\prime})^2 L =0$,
where primes here denote differentiation with respect to $w$. The effective energy-momentum tensor given in Eq. (\ref{GWT}) can then be written as
\be
\label{GWT2}
8 \pi \mathcal{T}_{ab} = \frac{2 \beta}{L^2} \left( \gamma^{\prime} L^2 \right)^{\prime} k_a k_b,
\ee
where $k^a=\sqrt{2} \delta^a_{\phantom{a} v}$ is the tangent vector to the single principal null direction. It is tempting at this point to compare this expression to the effective energy-momentum tensor for high-frequency gravitational waves found by Isaacson \cite{isaac1,isaac2}. In this appoach one considers fluctuations of amplitude $\epsilon \ll 1$, and imposes the condition that derivatives acting on wave-like fluctuations are of order $\epsilon^{-1}$. For the solution above this corresponds to taking $\gamma \sim O(\epsilon)$, so that $L=$constant$ + O(\epsilon)$. Eq. (\ref{GWT2}) then becomes
\be
8 \pi \mathcal{T}_{ab} = 2 \beta \gamma^{\prime \prime} k_a k_b + O(1).
\ee
This expression is a factor of $\epsilon^{-1}$ larger than the effective energy-momentum tensor found by Isaacson, which in the present case would be given by
\be
8 \pi \mathcal{T}_{ab}^{\rm Isaacson} = 2  \left(\gamma^{\prime}\right)^2 k_a k_b + O(\epsilon).
\ee
The effective energy-momentum tensor for wave-like gravitational fields that we have constructed above is therefore {\it not} the same as the one that appears on the right-hand side of the field equations in Isaacson's approach. This is not too surprising, as we have constructed our definition from the Weyl curvature, while Isaacson's definition is taken to be proportional to the Ricci curvature through the field equations.

\subsection{Temperature of Gravitational Fields}
\label{Temp}

In the discussion presented so far in this section we have calculated expressions for $T_{\rm grav} \dot{s}_{\rm grav}$ by appealing to the analogy between the effective energy-momentum tensor of the gravitational fields that we have constructed, and the actual energy-momentum tensor of matter fields. In order to go from these equations to a calculation of entropy specifically we now need to know the {``temperature'', $T_{\rm grav}$},
of the free gravitational fields. It is normally the case that, when considering the thermodynamics of ordinary matter fields, it is not possible to determine the temperature of a fluid without knowing something about the underlying microscopic physics. In the kinetic theory of gases, for example, one can identify temperature with the average kinetic energy of the gas molecules, but one cannot determine temperature from consideration of the macroscopic thermodynamic variables only. In this regard, it is not unreasonable to assume that a thermodynamic consideration of the free gravitational fields should be any different to that of standard thermodynamics: In order to know the temperature one needs to know something of the physics of the underlying microscopic theory.

For the temperature of gravitational fields we therefore appeal to the results of black hole thermodynamics \cite{bh}, and quantum field theory in curved spacetimes \cite{qft}. For our purposes we require a definition of temperature that is local (rather than being defined for horizons only), and that reproduces the expected results from semi-classical calculations in spacetimes such as Schwarzschild and de Sitter. We therefore take the temperature at any point in spacetime to be given by the following expression:
\be
\label{temp}
T_{\rm grav} = \frac{\vert u_{a ; b} l^a k^b \vert}{\pi} = \frac{\vert \dot{u}_a z^a + H + \sigma_{ab} z^a z^b \vert}{2 \pi},
\ee
where $l^a = (u^a-z^a)/\sqrt{2}$ and $k^a=(u^a+z^a)/\sqrt{2}$ are the real vectors in a complex null tetrad, $z^a$ is a spacelike unit vector aligned with the Weyl principal tetrad, and $H=\Theta/3$ is the isotropic Hubble rate. This expression reproduces the Hawking temperature \cite{bh}, the Unruh temperature \cite{unruh} and the temperature of de Sitter space \cite{gh} in the appropriate limits. One should note that Eq. (\ref{temp}) is an extra ingredient in our analysis, beyond the construction of $\mathcal{T}_{ab}$, and one could use alternative definitions as appropriate\footnote{This may, for example, be the case if one wishes to try and use perturbed Friedmann-Lema\^{i}tre-Robertson-Walker solutions as effective macroscopic descriptions of more complicated microscopic spacetimes. The underlying theory would then be Einstein's theory, and the temperature should presumably arise out of the consideration of averaging or coarse-graining procedures.}.

\section{The Entropy of Black Holes}
\label{Sec5}

The entropy of stationary black hole spacetimes is of obvious importance in discussing gravitational entropy, as they are so far the only spacetimes that allow an unambiguous definition of entropy in free gravitational fields. We will therefore consider them here, in the light of our previous discussion,
in order to investigate whether our proposal agrees with previously established results.

We consider the Schwarzschild geometry written in Gullstrand-Painlev\'{e} coordinates as
\be
ds^2 = -\left(1-\frac{2m}{r} \right)dt^2 -2 \sqrt{\frac{2m}{r}} dr dt +dr^2 +r^2 d\Omega^2,
\ee
where $m$ is the constant mass parameter. This coordinate system admits hypersurfaces of constant $t$ that intersect the horizon, and that have Euclidean geometry. For our unit vectors $u^a$ and $z^a$ we take
\begin{eqnarray}
&u_a = \left(0 ; \frac{1}{\sqrt{\vert 1- \frac{2 m}{r} \vert}} ,0,0 \right)\\
&z^a = \left( \frac{1}{\sqrt{\vert 1- \frac{2 m}{r} \vert}};0 ,0,0 \right),
\end{eqnarray}
such that in the region $r < 2m$ we have $u^a u_a=-1$ and $z^a z_a=1$. It is also the case that $u^a z_a =0$, and $z^a$ is orthogonal to the Euclidean hypersurfaces with $t=$constant. This choice is such that $u_a$ and $z^a$ specify a Weyl principal tetrad, as in Eq. (\ref{nullt}).

From Eqs. (\ref{rho}) and (\ref{temp}) we then have that the gravitational energy density and temperature is given at each point in the region $r<2m$ by
\begin{eqnarray}
&8 \pi \rho_{\rm grav}
= \alpha \frac{2m}{r^3} \\
&T_{\rm grav}= \frac{m}{2 \pi r^2 \sqrt{\vert 1- \frac{2 m}{r} \vert}}.
\end{eqnarray}
Now, Eq. (\ref{gibbsD}) cannot be directly applied to stationary spacetimes as strictly it involves changes in thermodynamic quantities over time. Following the methods used in \cite{bh}, we therefore instead choose to compare two different stationary black hole spacetimes that might represent the gravitational field before and after a small amount of mass is added to the black hole. Eq. (\ref{gibbsD}) then becomes
\be
\delta s_{\rm grav} = \frac{\delta (\rho_{\rm grav}v)}{T_{\rm grav}},
\ee
for an incremental increase in effective gravitational energy at a given $T_{\rm grav}$ along each of the curves $u^a$. Integrating over a volume $V$ on a hypersurface of constant $t$ then gives
\be
S_{\rm grav} = \int_V s_{\rm grav} = \int_V \frac{\rho_{\rm grav} v}{T_{\rm grav}} = \int_V \frac{\alpha r}{2} \sin \theta dr d\theta d\phi,
\ee
where we have taken $v=z^a \eta_{abcd} dx^b dx^c dx^d$ and set an arbitrary constant to zero. If $V$ is the region interior to the event horizon then this gives
\be
S_{\rm grav} = \alpha \frac{A_{\rm hor}}{4},
\ee
where $A_{\rm hor} = 4 \pi (2m)^2$ is the area of the horizon. This expression is identical to the Bekenstein-Hawking value if $\alpha = 1$, and therefore satisfies point {\bf E4}.

In order to consider the entropy in more general black hole spacetimes it will be necessary to include the entropy associated with the extra degrees of freedom involved. For the Reissner-–Nordstr\"{o}m solution this will mean including the energy density associated with the electromagnetic field, while in the Kerr solution it will require taking into account the energy associated with the rotation. Nevertheless, we see no reason in principle why the above approach could not be applied to these solutions. They are all Petrov type D, so the same equations should be valid. We leave this for future work.

\section{The Entropy of Large-Scale Structure in Cosmology}
\label{lss}

Let us now consider a spatially flat Robertson-Walker geometry with scalar perturbations in a longitudinal gauge, such that the line-element can be written
\be
\label{rw}
ds^2= a^2(\tau) \left[ -(1+2 \phi) d\tau^2 +(1-2 \phi) (dx^2+dy^2+dz^2) \right].
\ee
Our reference set of curves in this spacetime will be given by
\begin{eqnarray}
\label{urw}
&u^a = \left( \frac{1}{a}(1-\phi); u^i \right) \\
\label{zrw}
&z^a = \frac{1}{a \vert \nabla \phi \vert} \left( 0 ; \nabla_i \phi \right),
\end{eqnarray}
where $i,j$ are spatial indices, and higher-order terms have been discarded.

To lowest order in perturbations we then have that the space-time is silent, with $H_{ab}=0$. Meanwhile, the electric part of the Weyl tensor is given in tetrad components by
\be
\label{ab}
E_{AB}=e_A^{\phantom{A}a} e_B^{\phantom{B} b} E_{ab} = \phi_{.AB} +\sum_{C=1}^3 \phi_{.A} \gamma_{CAB}  \hspace{1cm} {\rm for} \hspace{1cm} \; A \neq B
\ee
where $\gamma_{ABC}=e_{A a ; b} e_{B}^{\phantom{B} a} e_{C}^{\phantom{C}b}$ are the Ricci rotation coefficients, and a dot denotes the tetrad component of a covariant derivative such that $\phi_{.AB}= e_B^{\phantom{B} b} (e_A^{\phantom{A}a} \phi_{;a})_{;b}$. By aligning ${\bf e}_3$ with $\nabla \phi$ we have
$\phi_{.1}=0=\phi_{.2}$ everywhere, and it can be deduced from Eq. (\ref{ab}) that in this case $E_{ab}$ is diagonalized with
$
E_{ab} = \sum_{A=1}^3 \lambda_A e_{Aa} e_{A b},
$
where $\lambda_A$ are the eigenvalues of this tensor. The Petrov type of the spacetime in Eq. (\ref{rw}) must therefore be either I, D or O, to lowest order in perturbations\footnote{What we mean by this is that the lowest order scalar perturbations in Eq. (\ref{rw}) lead to a Weyl tensor that is of one of these Petrov types, when all higher order terms are neglected.}. That the diagonal form of $E_{ab}$ should be maintained after spatial rotations means that $\lambda_1=\lambda_2$, so the spacetime is in fact type D (unless $\phi=$constant, in which case it is type O).

Aligning $z^a$ with ${\bf e}_3$, as in Eq. (\ref{zrw}), and using Eqs.  (\ref{rho}), (\ref{temp}) and (\ref{urw}), we find the gravitational energy density and temperature to be given to lowest order by
\begin{eqnarray}
&8 \pi \rho_{\rm grav}
=\alpha \vert \phi_{,\langle i j \rangle} z^i z^j \vert
= \alpha \frac{\left\vert (a^4 u_{\langle i , j \rangle}) \dot{\;} z^i z^j \right\vert}{a^3}\\
\label{trw}
&T_{\rm grav} = \frac{\vert H \vert}{2\pi},
\end{eqnarray}
where a dot denotes a derivative with respect to $\tau$, and angled brackets around indices denote that a quantity is tracefree.

After integrating over a spatial volume $V$, Eq. (\ref{gibbsD}) now gives the evolution of the gravitational entropy of this spacetime to lowest order as
\bea
\dot{S}_{\rm grav} &=& \frac{\alpha}{2 \vert H \vert} \int_V \Big( a^3 \sqrt{W/6} \dot{\Big)\;} dx dy dz\\
&=& \frac{\alpha}{4 \vert H \vert} \int_V \Big( \left\vert (a^4 u_{\langle i , j \rangle} \dot{)\;} z^i z^j \right\vert \dot{\Big)\;} dx dy dz,
\eea
where we have used $v=a^3 dx dy dz$. We take the spatial domain $V$ to be a comoving spatial volume, although one is free to make other choices. It can be seen that our choice of temperature in
Eq. (\ref{trw}) reproduces the expected Gibbons-Hawking temperature of de Sitter space \cite{gh}, as well as the temperature of the horizon in other homogeneous and isotropic spaces \cite{frwtemp}.

This formulation of gravitational entropy can be seen to be directly dependent on the shear of $u^i$, and on the anisotropy of the gravitational field through $\phi_{,\langle ij \rangle}$. It therefore explicitly satisfies point ${\bf E3}$. Factoring $\phi$ and $u^i$ into time-dependent and time-independent parts, as is usual in the study of perturbed Robertson-Walker geometries, it can be seen that the approximate time dependence of $S_{\rm grav}$ goes like
\be
\label{time}
S_{\rm grav} \sim
a^3 \tilde{u},
\ee
where $u_i = \tilde{u}(\tau) Q_i(x^j)$, and where $Q_i(x^j)$ is a harmonic function. Now for the monotonicity condition ${\bf E5}$ to be satisfied,
the quantity on the right-hand side of Eq. (\ref{time}) needs to be a monotonically increasing function
of time. For expanding dust dominated universes we note that Eq. (\ref{time}) gives
\be
S_{\rm grav} \sim \tau^5 \sim t^{5/3},
\ee
where $t=\int a(\tau) d\tau$ is the proper time of comoving observers. This certainly satisfies the monotonicity condition. In fact, the $S_{\rm grav}$ in Eq. (\ref{time}) will grow for any $\tilde{u}$ that decays more slowly than $\sim a^{-3}$.

\section{A Non-perturbative Inhomogeneous Example}
\label{Sec7}

It would be of interest to see whether the above measure of entropy is also applicable
in non-perturbative inhomogeneous settings. To this end, let us
consider the Lema\^{i}tre-Tolman-Bondi (LTB) solution, with line-element given by
\be
ds^2 = -dt^2 + \frac{R^{\prime 2} dr^2}{1-k(r)} +R^2 d\Omega^2,
\ee
where $R=R(r,t)$, and prime denotes partial differentiation with respect to $r$.  The field equations then reduce to
\be
\frac{\dot{R}^2}{R^2} = \frac{m(r)}{R^3} - \frac{k(r)}{R^2} \qquad {\rm and } \qquad
\rho = \frac{m^{\prime}}{8 \pi R^2 R^{\prime}},
\ee
where dots here denote differentiation with resect to $t$.
The free function $m(r)$ in this solution specifies a measure of the gravitational mass within a sphere of radius $r$, and in the FLRW limit reduces to $m \rightarrow 4 \pi R^3 \rho /3$. The LTB spacetime is of Petrov type D, unless it is FLRW.

If we now take our time-like and space-like unit vectors to be
\bea
u^a &=& \left(1; 0, 0 ,0 \right)\\
z^a &=& \left( 0; \frac{\sqrt{1-k(r)}}{R^{\prime}},0,0 \right),
\eea
then we obtain the following expressions for the effective energy density and temperature of the free gravitational fields
\bea
8\pi \rho_{\rm grav} &=& 2 \alpha \frac{\vert m-4 \pi R^3 \rho/3 \vert}{R^3}\\
T_{\rm grav} &=& \frac{1}{2\pi} \left\vert \frac{\dot{R}^{\prime}}{R^{\prime}} \right\vert.
\eea
The effective energy density grows as the departure from FLRW increases, weighted by a factor of $R^3$,
and the effective temperature is simply proportional to the expansion in the radial direction.

In a given LTB spacetime, entropy will then increase as long as the expansion weighted value
of $\rho_{\rm grav}$ increases.
For a given LTB solution, we therefore have an arrow of time that associates low $\rho_{\rm grav}$ with early times, and high $\rho_{\rm grav}$ with late times. This provides us with a way of specifying which direction is ``the future'' for any given solution (independent of whether the coordinate $t$ happens to be increasing in that direction or not, or whether the spacetime is expanding or not) \cite{Ell13}. It also supports our intuitive understanding of gravitational entropy in cosmology, in which the low entropy state of the early universe was close to homogeneous and isotropic, and in which the late universe is inhomogeneous and anisotropic.

What this means for an example such as a time-reversal-symmetric recollapsing LTB solution that starts off close to homogeneous, then evolves towards inhomogeneity at its maximum of expansion, before recollapsing to a near homogeneous final state, remains to be seen. It seems safe to conjecture that solutions of this type will always be problematic for any sensible definition of gravitational entropy, however, as time reversal symmetry about the maximum of expansion will necessarily mean that any notion of entropy based on the geometry of spacetime will not be monotonic throughout its entire evolution. An example such as this would appear to have an entropic arrow of time that points towards the maximum of expansion from both sides, so that the ``future'' (defined in this way) is always in the direction of positive cosmological expansion.

\section{Discussion}
\label{Sec8}

Motivated by thermodynamical considerations, we have proposed a
measure of gravitational entropy based on the square-root of
the Bel-Robinson tensor. A key feature of this measure is
its non-negativeness, which is a fundamental requirement
for any measure of entropy, and is in contrast to other gravitational
entropy measures previously proposed. We have applied our measure to a number of
examples, including cosmological ones, and have found that the specific form of this
measure will depend on whether the gravitational field is
Coulomb-like or wave-like. 

In constructing this measure of gravitational entropy we have assumed that
the square-root of the Bel-Robinson tensor can be taken to be the effective energy-momentum
tensor of free gravitational fields, that the energy of this effective fluid is conserved in vacuum, 
that the temperature of gravitational fields can be
determined from semi-classical considerations, and that the fundamental equation of
thermodynamics is applicable. We have also assumed that the entropy of the universe
at any given time is given by integrating the entropy density over a space-like hypersurface.
Under these assumptions our definition of 
entropy in free gravitational fields is unique. For the square-root to exist as a
unique factorisation of the Bel-Robinson, however, requires the spacetime to contain
gravitational fields that are only Coulomb-like or only wave-like. The factorisations of the
Bel-Robinson tensor that are possible when both of these types of field are present
are more complicated, and are not unique. Further study is therefore necessary to
extend this definition to the case of general spacetimes.

In the case of the Schwarzschild black hole,
our entropy measure reduces to the usual Bekenstein-Hawking value.
For scalar perturbations of a Robertson-Walker geometry we find
that our measure evolves like the Hubble weighted anisotropy of
the free gravitational field. As a result it increases as structure
formation occurs, as is expected of a sensible measure.
We have also applied our measure to the non-perturbative case of
LTB models and found conditions under which the entropy increases.
These examples, ranging from black holes to perturbed FLRW and
exact inhomogeneous LTB, provide encouraging evidence that our proposed measure
has the potential of accounting for gravitational entropy
in a range of cosmological settings of interest. However, it is only in the black hole
case that there is and obvious link to the holographic principle \cite{holo}.

Finally, we note that the definition of gravitational entropy we have considered 
in this paper is only valid for General Relativity. In other theories of gravity the Bel-Robinson
may not be the appropriate choice to describe the super-energy-momentum of
free gravitational fields, and a Bel-Robinson-like tensor may not exist at all \cite{deser}.

\ack

We are grateful to A. Coley, R. Maartens, R. Penrose and J.-P. Uzan for helpful discussions and suggestions. TC acknowledges the support of the STFC, and is grateful to the Department of Mathematics and Statistics at Dalhousie University for hospitality while some of this work was carried out. TC and RT are both grateful to the Department of Mathematics and Applied Mathematics at the University of Cape Town for hospitality.

\end{document}